\documentclass[twoside]{article}
\usepackage{iwce}
\usepackage{graphicx}
\begin{document}

\title{Modeling of Transport through Submicron Semiconductor
  Structures: A Direct Solution of the Coupled Poisson-Boltzmann Equations}

\IWCEauthorsFirst{D. CSONTOS AND S.E. ULLOA}
\setHeadings{Csontos}{A Direct solution to the Coupled
  Poisson-Boltzmann Equations} 
\IWCEaddressFirst{Department of Physics and Astronomy, and Nanoscale and
  Quantum Phenomena Institute, Ohio
University, Athens, Ohio 45701, USA} 
\email{csontos@phy.ohiou.edu}

\preparetitle

\begin{IWCEabstract}
We report on a computational approach based on the self-consistent
solution of the steady-state Boltzmann transport equation coupled with
the Poisson equation for the study of inhomogeneous transport in deep
submicron semiconductor structures. The nonlinear, coupled
Poisson-Boltzmann system is solved numerically using finite difference
and relaxation methods. We demonstrate our method by calculating the
high-temperature transport characteristics of an inhomogeneously doped
submicron GaAs structure where the large and inhomogeneous built-in
fields produce an interesting fine structure in the high-energy tail
of the electron velocity distribution, which in general is very far from
a drifted-Maxwellian picture.  
\end{IWCEabstract}

\IWCEkeywords{Boltzmann equation, electron distribution function,
  submicron devices, upwind method}

\begin{multicols}{2}
\intro{Introduction}
The carrier dynamics in submicron structures is far from thermal
equilibrium due to strong and rapidly varying external and built-in
electric fields. Hot electron and ballistic effects dominate the
transport characteristics and the electron velocity distribution
function in such systems is far from a drifted-Maxwellian
description. In order to fully take into account the  
nonequilibrium nature of the transport, a full solution of the
semiclassical Boltzmann transport equation (BTE) is required. Although
the Monte Carlo method has been very popular for the solution of the
BTE in semiconductor device simulation \cite{jacoboni}, 
several works \cite{barangerPRB87}-\cite{majoranaCOMPEL04} have
recently solved the BTE by direct methods,  
thus allowing noise-free spatial and temporal resolution of the
electron distribution function, which in the Monte Carlo method may be
difficult to obtain due to the statistical nature of the approach. In
this paper, we present a straight-forward approach to calculate the 
electron distribution function, $f(x,v)$, for submicron inhomogeneous
semiconductor 
structures by solving the steady-state BTE self-consistently with the
Poisson equation. We solve the strictly
two-dimensional (2D) BTE (one dimension corresponding to position and
one to velocity) and treat
scattering within the relaxation time 
approximation (RTA) where each individual scattering mechanism is
represented by a characteristic scattering rate that can be derived
from quantum mechanical scattering theory. We demonstrate our approach for
submicron, inhomogeneously doped structures and discuss the general
nonequilibrium transport characteristics.    

\section{Basic equations}
\noindent
The Boltzmann equation describes the dynamics of the semiclassical
distribution function, $f({\bf r}, {\bf v}, t)$, under the influence
of electric and magnetic fields, as well as different scattering
processes. In the absence of a magnetic field, the 2D phase-space,
steady-state BTE in the RTA is written according to: 
\begin{equation}
 -\frac{eE(x)}{m^{\ast}}\frac{\partial f(x,v)}{\partial
 v}+ v\frac{\partial 
 f(x,v)}{\partial x}=-\frac{f(x,v)-f_{LE}(x,v)}{\tau(\varepsilon)}~,
\label{bte}
\end{equation}
where $m^{\ast}$ is the electron effective mass in the parabolic band
approximation, and $f_{LE}(x,v)$ is a local equilibrium distribution
function appropriate to a local density, applied field and equilibrium
lattice temperature, $T_{0}$, to which the distribution function
$f(x,v)$ relaxes at a relaxation rate $\tau(\varepsilon)^{-1}$. As the
local equilibrium function, we choose in the following a
Maxwell-Boltzmann (MB) distribution at $T_{0}$, normalized to the
local density $n(x)$
\begin{equation}
f_{LE}(x,v)=n(x)\left [ \frac{m^{\ast}}{2\pi kT_{0}} \right ]^{1/2}
  e ^{-\frac{m^{\ast} v^{2}}{2k_{B}T_{0}}}~.
\label{mb}
\end{equation}

The inhomogeneous electric field, $E(x)$, in the BTE, originating from
the spatially dependent electron and doping densities, $n(x)$ and
$N_{D}(x)$, is obtained from the Poisson equation  
\begin{equation}
\frac{d ^{2} \phi}{d x^{2}}  = -\frac{dE}{dx}= -e \frac{N_{D}(x) -
  n(x)}{\epsilon \epsilon_{0}} = -\rho(x),
\label{poisson}
\end{equation}
where $\epsilon$ is the static dielectric constant. Since the electron
density is related to the distribution function by
\begin{equation}
n(x)=\int^{\infty}_{-\infty} f(x,v)dv~,
\label{density}
\end{equation}
the Poisson and Boltzmann equations constitute a coupled, nonlinear set of
equations, and thus, Eqs. (\ref{bte}-\ref{density}) have to be solved
self-consistently. 

\section{Numerical procedure}
\noindent
The numerical procedure consists, in short, of initializing the system
parameters, discretizing
Eqs. (\ref{bte}-\ref{density}) on a 2D grid in phase-space, performing
the self-consistent Poisson-Boltzmann loop and, upon convergence,
calculate and output the electron distribution function, electric
field and the desired moments of the BTE. In the calculations, after
initialization, we rescale the system parameters and the equations
according to 
\begin{equation}
x^{\prime}=x/L_{D},~v^{\prime}=v\tau/L_{D},
\label{scaling}
\end{equation}
where $L_{D}=\sqrt{\epsilon \epsilon_{0}k_{B}T_{0}/e^{2}N}$ is the
Debye length, $N=\max[N_{D}(x)]$ and $\tau$ is a characteristic
scattering time. The choice of grid size and resolution depends to a large
extent on the system parameters and the electrostatics present in the
device. In order to reproduce details due to strong and rapidly
varying electric fields, we choose the spatial grid step size to be
smaller than the Debye length, $L_{D}$, defined above. In velocity
space, on the other hand, the discrete grid step size needs to be
small enough to resolve fine structure in the distribution function,
as well as give accurate results for the moments of the BTE. In
addition, the grid needs to be large enough, in velocity, in order to
capture the full information in the high-energy tail of the
distribution function, and in position, in order to damp out the
effects of the contact boundaries.     

The Poisson and Boltzmann equations are solved by finite difference
and iterative relaxation methods \cite{numrec}. For the Poisson
equation (\ref{poisson}), we use forward and backward Euler
differences according to 
\begin{equation}
L^{+}_{x}L^{-}_{x}\phi_{j}=\frac{\phi_{j+1}-2\phi_{j}+
  \phi_{j-1}}{(\delta x)^{2}}= -\rho_{j}~,
\label{poissondifference}
\end{equation}
where $L^{+}_{x}\phi(x)=(\phi_{j+1}-\phi_{j})/\delta x$ and
$L^{-}_{x}\phi(x)=(\phi_{j}-\phi_{j-1})/\delta x$ denote forward and backward
Euler steps, respectively. The resulting matrix equation is solved
iteratively using successive overrelaxation (SOR) \cite{numrec}.

For the solution of the BTE, we adopt an upwind finite difference scheme
\cite{fatemiJCP93} which amounts to the following discretization of
the partial derivatives in Eq. (\ref{bte}):
\begin{eqnarray}
\frac{\partial f}{\partial v} & = &
L_{v}^{+[-]}f(x,v)~~E(x)>0~[E(x)\leq 0] \\
\frac{\partial f}{\partial x} & = & L_{x}^{+[-]}f(x,v)~~v<0~[v\geq 0]~.
\label{btedifference}
\end{eqnarray}
As for the Poisson equation, we use SOR to solve the matrix equation
resulting from the discretization of Eq. (\ref{bte}). 

For the boundary conditions of the Poisson-Boltzmann equations we
adopt the following: For the potential, the values at the system
boundaries, denoted (l)eft and (r)ight are fixed to
$\phi(x_{l})=U_{0}$ and $\phi(x_{r})=0$, respectively,
corresponding to an externally applied voltage $U_{0}$. The electron
density is allowed to fluctuate freely around the boundaries, subject
to the condition of global charge neutrality, which is enforced between
each successive iteration in the self-consistent Poisson-Boltzmann
loop. We choose the size of the highly-doped contacts to be large
enough such that the electron density and the electric field deep
inside the contacts is constant. 

For the electron distribution
function four boundary conditions can be defined in the 2D phase-space. At
the velocity cut-off in phase-space, we choose
$f(x,v_{max})=f(x,-v_{max})= f_{LE}(x,v)$ which is reasonable since we
assume $v_{max}\geq 30k_{B}T_{0}$ in the calculations. At such high
velocities, the 
electron population is negligible and of the same order as the local
equilibrium distribution $f_{LE}(x,v)$. At the contact boundaries, we
assume that the electric field is low and constant (as verified in the
calculations), and thus, the homogeneous solution to the BTE in the
linear response regime of transport applies. Hence,
\begin{equation}
f(x_{i},v)=f_{LE}(x_{i},v)[1-vE(x_{i})\tau(\varepsilon)/k_{B}T_{0}],
\end{equation}
where $i=l,r$. The iterative Poisson-Boltzmann loop consists of an
updating procedure 
for the electric field, electron distribution function and electron
density using Eqs. (\ref{poisson}, \ref{bte}, \ref{density}), until
convergence. The convergence criterion is determined and checked in
terms of the 
evolution of the $L_{2}$ norm of the potential and density variations
between subsequent iterations. Typically, the results are converged
when the $L_{2}$ norms for the potential and density are on the order
of $10^{-3}$ of the original values. Between subsequent iterations, we
employ linear mixing in the electron density, according to
\begin{equation}
n^{\prime}(x)=(1-\alpha) n^{old}(x)+\alpha n^{new}(x)~,
\end{equation}
where $n^{old}(x)$ is the input density to the Poisson solver,
$n^{new}(x)$ is the new density obtained from the solution of the BTE
using the new electric field obtained from the Poisson solver, and
$n^{\prime}(x)$ is the final density that is used as an input to the
next iteration in the Poisson-Boltzmann loop. The convergence and
stability of the self-consistent loop are strongly dependent on the
system parameters and the nonequilibrium nature of the electronic
system. If the system is strongly out of equilibrium, displaying large
variations and strengths of the electric field, the mixing parameters
$\alpha$ may have to be chosen as small as a few percent, thus
affecting the overall runtime. Furthermore, for highly doped
structures, the convergence is slower, partly due to the required
small grid size in position due to the small Debye length, but also
due to the slow convergence in the SOR procedure  in the BTE, where the
stability of the numerical scheme is given in terms of a
Courant-Friedrich-Levy type condition \cite{numrec}. Still, the
computational demands for the calculations reported in this paper are
modest.
\section{Numerical results}
\noindent
In the following, we demonstrate our numerical approach with
calculations of the transport characteristics of a model GaAs
$n^{+}-n^{-}-n^{+}-n^{-}-n^{+}$ structure with the doping
densities $n^{+}$=10$^{23}$ m$^{-3}$ and $n^{-}$=10$^{19}$
m$^{-3}$. In order to highlight the effects of inhomogeneities and
scattering while keeping the nature of the scattering structureless,
we use a constant scattering time $\tau=2.5\cdot 10^{-13}$ s, which
corresponds to realistic mobilities of GaAs at room temperature for
which the calculations have been performed. The central
$n^{-}-n^{+}-n^{-}$ region has the dimensions 200/200/200 nm, whereas 
the contacts are 1 $\mu$m long. 

\begin{IWCEfigure}
  \centering
\includegraphics*[width=0.7\textwidth]{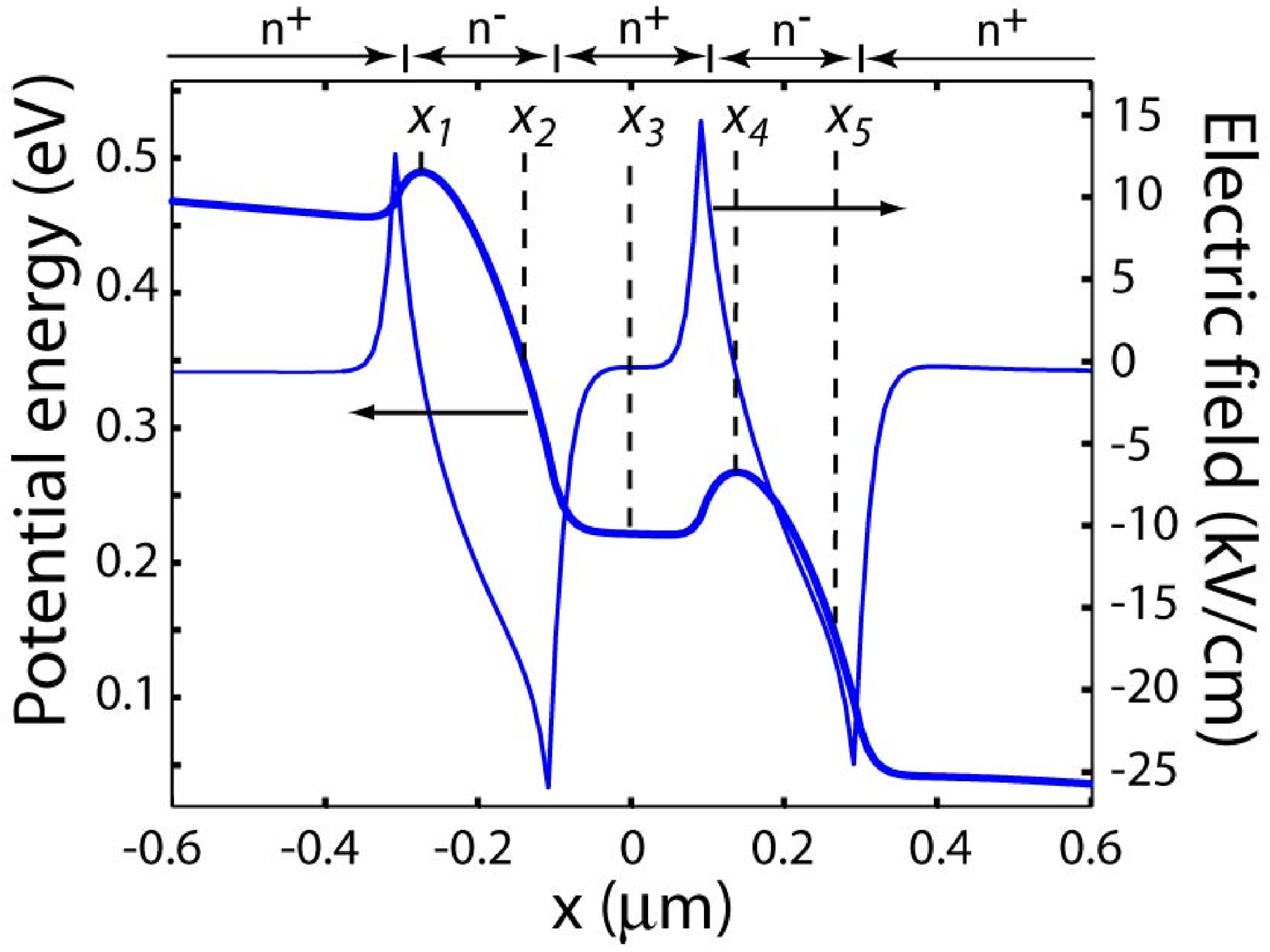}
  \caption{Electron potential energy and electric field in the central
  region of the $n^{+}-n^{-}-n^{+}-n^{-}-n^{+}$ system with
  dimensions described in the text. The corresponding current density
  is $j(x)=-5.4\cdot 10^{4}$ A/cm$^{2}$.}
  \label{fig1}
\end{IWCEfigure}

In Fig. \ref{fig1} we show the electric field and potential energy
around the central region of the system described above, subject to an
applied bias voltage $V_{b}=-0.5$ V. Due to the charge imbalance,
electrons diffuse towards the lightly doped regions, where potential
barriers are formed and, correspondingly, a large and inhomogeneous
electric field on the order of 10 kV/cm is formed, even in
the absence of an external applied voltage. As a finite voltage is
applied to the device, the majority of the potential drop occurs
over the submicron central region, giving rise to a strongly
inhomogeneous field distribution, in contrast to the $n^{+}$ contact
regions, where the field in comparison is very low and constant.

The electron velocity distribution in the central region of the
structure is shown in Fig. \ref{fig2}(a), for five specific spatial
points as depicted in Fig. \ref{fig1}. Figure \ref{fig2}(b) shows a
contour plot of the full spatial dependence of $f(x,v)$ in that
region. It is clear that the inhomogeneous electric
field gives rise to a strong spatial dependence of the velocity
distribution function along the direction of transport, and that the
distribution function in the central region is very far from thermal
equilibrium. 

In the outermost highly doped $n^{+}$ regions, where the field is low
and constant the distribution is simply a shifted Maxwellian. In the
lightly doped $n^{-}$ regions on the other hand, the 
velocity distribution is highly asymmetric and develops a narrow peak
that rapidly shifts toward higher velocity along the direction of 
transport. This peak contains quasi-ballistic electrons which are
accelerated by the strong electric field in the central region, and
thus, have a considerably larger average velocity compared to the
electrons in the contacts. Close to the potential barrier the
distribution function is suppressed at low velocities due to the
skimming of the distribution of incoming electrons, as well as the
restriction of drain induced electron flow with $v<0$ due to the potential
barrier. 

However, the low-velocity contribution to the distribution
function gradually increases away from the barrier, as thermionically
injected electrons gradually are thermalized and the lower effective
barrier height allows electrons from the $n^{+}$ regions to penetrate
the lightly-doped region. Thus, the total distribution function
consists of a quasi-ballistic, high-velocity and a diffusive,
low-velocity contributions, which gives the total distribution function
a highly non-Maxwellian broad and asymmetric shape. Furthermore, the
presence of the two barriers creates an additional quasi-ballistic
structure in the high-energy tail of the distribution function in the
second $n^{-}$ region, as electrons that traverse the intermediate
$n^{+}$ region ballistically get an additional acceleration toward
higher velocities by the electric field in the second $n^{-}$ region,
thus creating two high-velocity electron beams. These features
emphasize the highly nonequilibrium nature of the electron transport
in these type of systems and demonstrate that our method is capable of
taking them fully into account.

\begin{IWCEfigure}
\centering
\includegraphics*[width=0.7\textwidth]{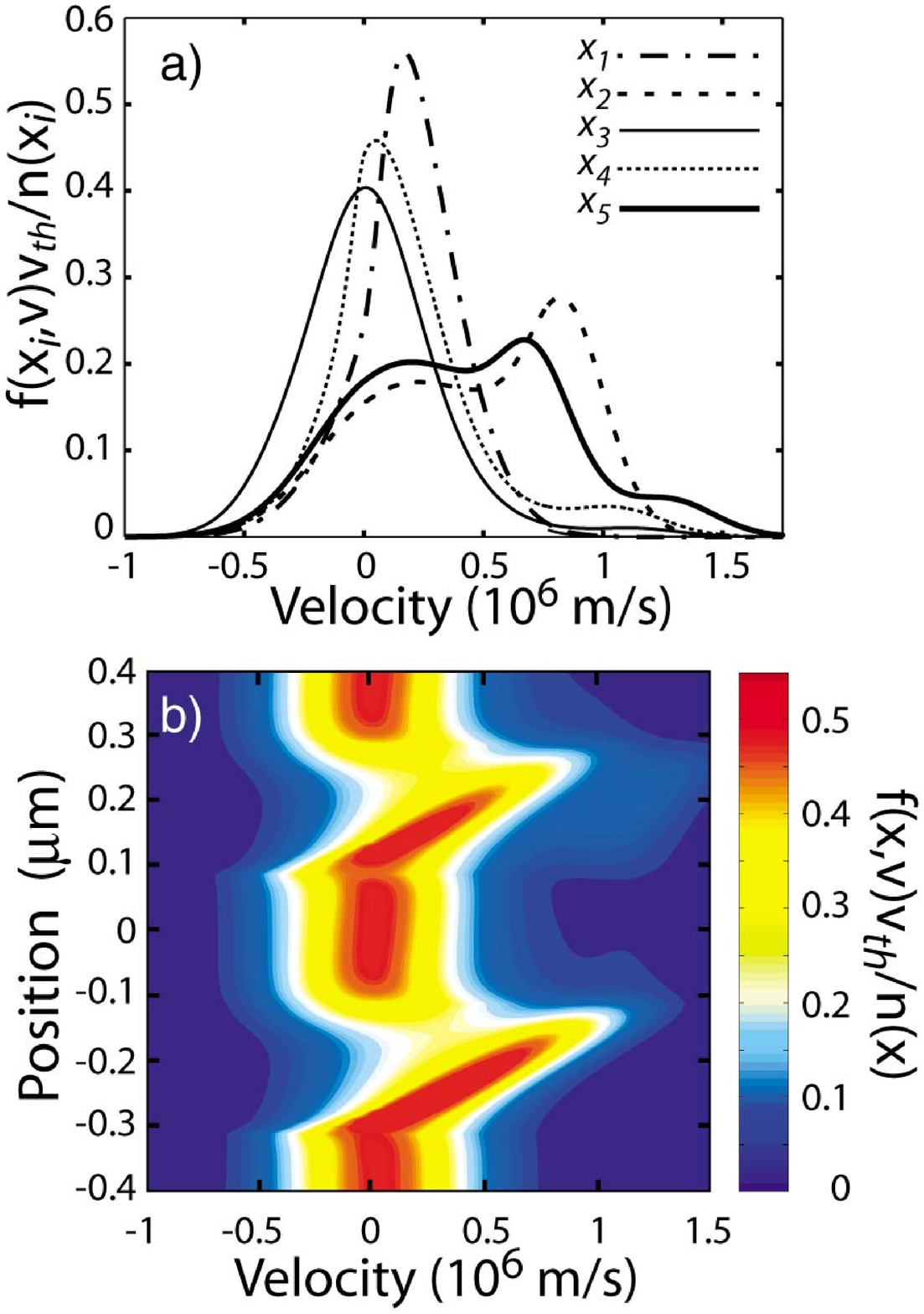}
  \caption{(Color online) Normalized electron velocity distribution
    function $f(x,v)v_{th} /n(x)$, where $v_{th}$ is the thermal
    velocity, at (a) five different points in space corresponding to
    the points depicted in Fig. \ref{fig1}. (b) Contour plot of
    $f(x,v)v_{th}/n(x)$.} 
  \label{fig2}
\end{IWCEfigure}

\section{Conclusions}
\noindent
We have presented a numerical method for the
solution of the steady-state, coupled Poisson-Boltzmann equations for
the study of inhomogeneous, submicron semiconductor structures and
demonstrated our approach on a submicron GaAs structure with strong
built-in electric fields. We have shown that our method is
capable of taking into account the strong nonequilibrium transport
properties that arise in such systems due to the presence of very
large and inhomogeneous electric fields, and that interesting
structure is present in the high-energy tail of the distribution
function, caused by quasi-ballistic electrons.

~\\
\noindent
{\bf Acknowledgments} \\
~\\
\noindent
This work was supported by the Indiana 21st Century Research and
Technology Fund.

\end{multicols}
\end{document}